\documentclass[twocolumn,showpacs,prl,superscriptaddress,tighten]{revtex4}

\usepackage{graphicx}
\usepackage{dcolumn}
\usepackage{bm}

\begin{document}

\preprint{PRL}

\title{The symmetry of the superconducting order parameter in PuCoGa$_5$ \\}

\author{G.D.~Morris}
\affiliation{Los Alamos National Laboratory, K764, Los Alamos, New Mexico 87545 USA.}
\author{R.H.~Heffner}
\affiliation{Los Alamos National Laboratory, K764, Los Alamos, New Mexico 87545 USA.}
\author{E.D.~Bauer}
\affiliation{Los Alamos National Laboratory, K764, Los Alamos, New Mexico 87545 USA.}
\author{L.A~Morales}
\affiliation{Los Alamos National Laboratory, K764, Los Alamos, New Mexico 87545 USA.}
\author{J.L.~Sarrao}
\affiliation{Los Alamos National Laboratory, K764, Los Alamos, New Mexico 87545 USA.}
\author{M.J.~Fluss}
\affiliation{Lawrence Livermore National Laboratory, P.O. Box 808, Livermore, California 94550 USA}
\author{D.E.~MacLaughlin}
\affiliation{Department of Physics, University of California, Riverside, California 92521 USA.}
\author{L.~Shu}
\affiliation{Department of Physics, University of California, Riverside, California 92521 USA.}
\author{J.E.~Anderson}
\affiliation{Department of Physics, University of California, Riverside, California 92521 USA.}
\date{\today}

\begin{abstract}

The symmetry of the superconducting order parameter in single-crystalline PuCoGa$_5$ ($T_{\rm c} = 18.5$~K) is investigated via  zero- and transverse- field muon spin relaxation ($\mu$SR) measurements,  probing the possible existence of orbital and/or spin moments (time reversal-symmetry violation TRV) associated with the superconducting phase and the in-plane magnetic-field penetration depth $\lambda(T)$ in the mixed state, respectively. We find no evidence for TRV, and show that the superfluid density, or alternatively, $\Delta\lambda(T) = \lambda(T) - \lambda(0)$, are $\propto T$ for $T/T_{\rm c} \leq 0.5$. Taken together these measurements are consistent with an even-parity (pseudo-spin singlet), d-wave pairing state.

\end{abstract}

\pacs{76.75.+i, 74.25.-q, 71.27.+a}
\maketitle

\typeout{PACS1}
\typeout{PACS2}
\typeout{PACS3}

Because  f-electrons may exist at the boundary between localization and itinerancy, f-electron compounds present an extreme challenge for many-body physics.  For example, Pu metal sits midway between `itinerant' Np and `localized' Am \cite{Pu-metal}, while  some theories for the $\delta$-phase of Pu postulate that  only one out of five of its 5-f electrons are delocalized \cite{Pu-theory}.   Delocalization is a key ingredient for heavy fermion behavior, which arises from strong f-electron conduction-electron hybridization.   A new class of tetragonal heavy-fermion compounds Ce$_n$MIn$_{3n+2}$, with $n = 1,2$,  has been under considerable investigation for the wide variety of magnetic and unconventional superconducting behaviors which it exhibits  \cite{115}.  The superconductivity in these compounds  occurs below $\sim 2$~K,  arising out of a normal state in close proximity to an antiferromagnetic quantum critical point and possessing a large Sommerfeld constant ($\gamma = 0.7$ J/mol-K$^2$ in CeIrIn$_5$).  The recent discovery \cite{Sarrao-2002} of superconductivity with an order-of-magnitude higher critical temperature in the same structural class of materials, but possessing Pu instead of Ce, is consequently both exciting and significant.  PuCoGa$_5$ becomes superconducting below $T_c = 18.5$~K from a normal state with a relatively modest $\gamma = 77$ mJ/mol-K$^2$  \cite{Sarrao-2002}. Photoemission suggests the f-electrons in PuCoGa$_5$ are between localized and itinerant \cite{Joyce}.

The superconducting pairing interaction in heavy fermion materials is believed to arise from antiferromagnetic spin fluctuations, and a similar pairing mechanism has been postulated for the light-mass cuprate superconductors \cite{Moriya}. This led to the speculation that PuCoGa$_5$ might be a bridging material between heavy fermions and the cuprates \cite{Sarrao-2002}.  Recent electronic structure calculations \cite{PuCoGa5-theory} for PuCoGa$_5$ show 2-dimensional Fermi surfaces which could support spin-fluctuation-mediated superconductivity and, hence, d-wave pairing \cite{Moriya}. However,  key measurements to confront these ideas are lacking; hence, tests of the symmetry of the superconducting order parameter in PuCoGa$_5$ are of supreme importance. 

This Letter reports muon spin relaxation experiments ($\mu$SR) in single-crystal PuCoGa$_5$ designed to explore two aspects of its superconductivity: (1) a possible violation of time-reversal symmetry (TRV), indicated by small, spontaneous magnetic fields 
below $T_c$,  and (2)  the temperature dependence and magnitude of its superconducting penetration depth $\lambda$. To date no direct spectroscopic probes of these quantities have been reported.
In these experiments spin-polarized positive muons ($\mu^+, S=1/2$) are stopped in the sample and precess in the local magnetic field ${\bf B}$ until they 
undergo a parity-violating weak decay
$\mu^+ \rightarrow \rm{e}^+ + \nu_{\rm e} + \bar{\nu_{\mu}}$ (lifetime $\tau_{\mu}$=2.2$\mu$s). The time-dependence of the polarization, monitored via the positron decay anisotropy, yields information about the local field inhomogeneity and fluctuation spectrum
\cite{muSR-reviews}.

Samples  were grown from excess Ga flux  \cite{Sarrao-2002}, forming flat plates with the $c$-axis
normal to the crystal face.
Two crystals measuring $\sim 5\times$6mm$^2$ in total area and $\sim$1/2mm
thick were encapsulated in a 70$\mu$m thick Kapton coating, and then attached  and sealed
under He atmosphere inside a titanium cell having a 50$\mu$m Ti-foil beam window and
attached to a continuous-flow He cold-finger cryostat. The dual encapsulation was undertaken to prevent possible radioactive contamination.

Conventional time-differential $\mu$SR experiments were performed
at the M20 channel at TRIUMF, Vancouver, Canada, only 3 weeks after the crystals were prepared.
Thus, there was no measurable degradation of $T_c$ due to radiation damage from Pu decay (half-life $\sim 2.4 \times 10^4$ y). Surface muons (momenta $\sim 29$ MeV/c) were implanted into the sample with their spins perpendicular to their momenta, in the plane of the sample face.  Approximately 1/3 of the muons stopped in the sample,
the remaining fraction  in the Ti sample holder.  A negligible fraction stopped in the Ti
window or the Kapton coating. The background signal obtained from an empty Ti
holder in either zero applied field (ZF) or $600$ Oe  applied transverse to the muon spin (TF) was well characterized  by a Gaussian relaxation function $G_{\rm Ti}(t) = \exp(-1/2\sigma_{\rm Ti}^2t^2$) with $\sigma_{\rm Ti} \approx 0.03 \mu$s$^{-1}$. For ZF experiments the residual field was reduced to $\leq 10$ mOe using trim coils.

\begin{figure}
\centerline{\includegraphics[width=\columnwidth]{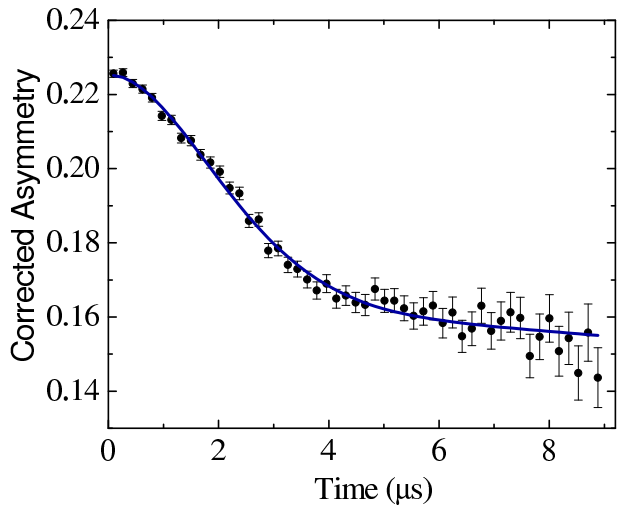}}
\caption{
A representative zero field spin relaxation spectrum at $T$=4~K.
The solid line is a fit to a dynamic Gaussian Kubo-Toyabe relaxation
function and a background from muons stopped in the Ti cell.}
\label{ZF}
\end{figure}

\begin{figure}
\centerline{\includegraphics[width=\columnwidth]{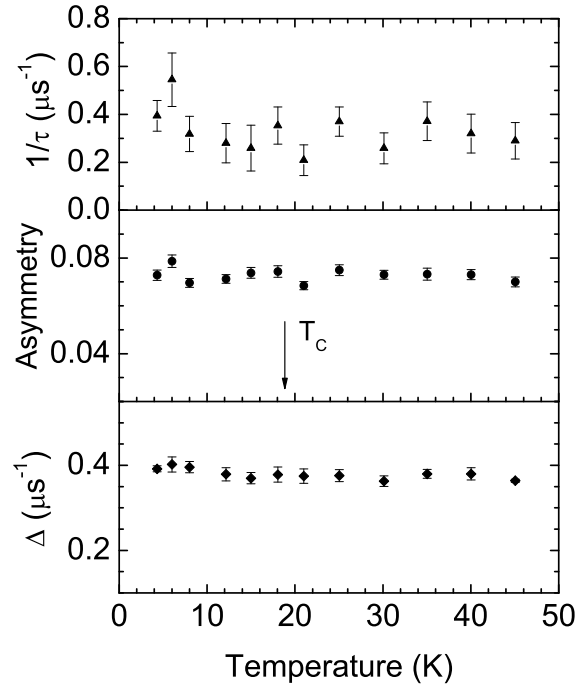}}
\caption{
Temperature dependencies of inverse correlation time $1/\tau$, amplitude and 
field width $\Delta$ obtained from fitting the ZF spectra to a dynamic Gaussian Kubo-Toyabe
relaxation function.}
\label{ZF-params}
\end{figure}

Figure~(\ref{ZF}) shows a representative ZF spin relaxation spectrum at $T$=4K,
corrected for differences in the positron counters'
efficiencies and solid angles.
All of the ZF spectra -- above and below $T_{\rm c}$ -- were very similar.
Between $T = 3-45$~K the overall relaxation is well described by the sum of $G_{\rm Ti}(t)$ 
and a \textit{dynamic} Gaussian Kubo-Toyabe
spin relaxation function \cite{muSR-reviews} $G_{KT}(t,\Delta,\tau)$, with  static rate $\Delta$ due to a distribution of Co and Ga nuclear
dipolar fields and dynamic rate $\tau$.  The latter is produced by slow 
field dynamics and/or muon motion, as also seen in CeCoIn$_5$ and CeIrIn$_5$ \cite{Higemoto}. In PuCoGa$_5$ these fluctuations   are sufficient to suppress the long-time tail of $G_{KT}$ (which approaches $1/3A_0G_{KT}(t=0)$ when $\tau = \infty$), but only slightly affect the initial Gaussian relaxation.
The temperature dependencies of $\tau^{-1}$, relaxing amplitude and rate $\Delta$
are shown in Fig.~\ref{ZF-params}.  No temperature dependence is seen between 
$T=3 - 45$ K.

Transverse field $\mu$SR locally probes the magnetic field distribution 
in the vortex state of superconductors \cite{Sonier}, which is characterized by $\lambda$
and the coherence length $\xi_0$, estimated to be $\xi_0 \sim 2.1$ nm from
the temperature dependence of the upper critical field $H_{\rm c2}$ near $T_{\rm c}$ \cite{Sarrao-2002}. 
Field-cooled TF measurements were carried out in $H_0 = 600$ Oe field applied parallel to 
the $c$-axis. This field is at least twice the lower critical field $H_{\rm c1}
\approx 300$ Oe reported previously \cite{Sarrao-2002} and $4-5 \times$ that implied by our
measurements of $\lambda(0)$ ($H_{\rm c1} \propto \ln(\lambda/\xi_0)/\lambda^2$), as discussed below.  Precession spectra were  fit to the sum of two terms, corresponding to muons
stopping in the sample and Ti cell, respectively:
\begin{eqnarray}
A_0 G_{\rm z}(t) &=& A \cos(\omega t + \phi) P(t) \nonumber \\
&+& A_{\rm Ti} \cos(\omega_{\rm Ti}t + \phi)\exp(-\sigma_{\rm Ti}^2 t^2/2).
\label{asym}
\end{eqnarray}

\begin{figure}
\centerline{\includegraphics[width=\columnwidth]{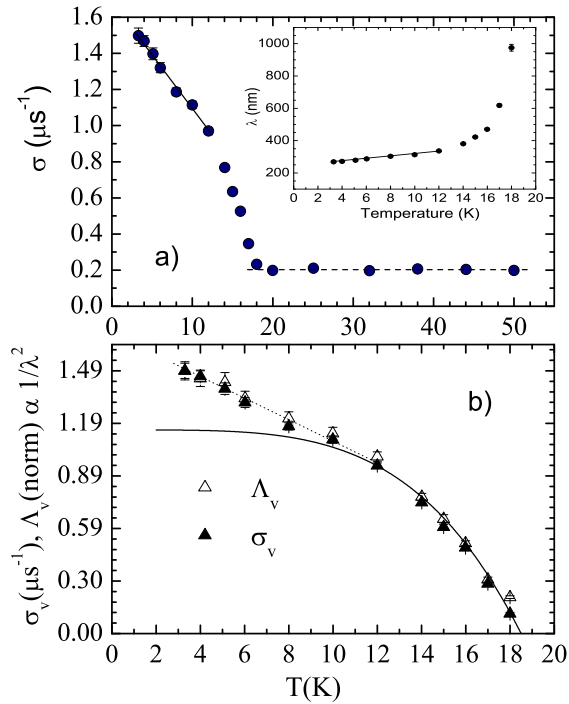}}
\caption{
a)  Temperature dependence of $\mu$SR Gaussian TF rate $\sigma(T)$. (See Eq. (1) and discussion in text.)  The average baseline
$\sigma_{\rm n}$ above $T_{\rm c}$ is indicated by the dashed line; the solid straight line shows $\sigma \sim T$ below $12$~K.  The inset shows $\lambda(T)$ obtained from the linewidth $\sigma_{\rm v}(T)$ and
Eq.~(\ref{eq:Brandt}). The solid line is a linear fit between $3 - 12$~K.  b)  Rates from Gaussian ($\sigma_{\rm v}$) and exponential ($\Lambda_{\rm v}$ - normalized to $\sigma_{\rm v}$ at $3.3$~K) forms for $P(t)$ in Eq. (1) have the same temperature dependence below $T_{\rm c}$.  The curve  is the result for the two-fluid model (s-wave), with $T_{\rm c} =18.5$~K and $\sigma_{\rm v}=1.15 \mu$s$^{-1}$, determined to fit the data for $T \geq 12$~K. The straight dotted line illustrates $\sigma_{\rm v}(T),\Lambda_{\rm v}(T) \propto T$ for $T \leq 12$~K.
}
\label{TF-2lw}
\end{figure}

The temperature dependence of $\sigma$ for a Gaussian $P(t) = \exp(-\sigma^2 t^2/2)$ is shown in Fig.~\ref{TF-2lw}a.  One sees that $\sigma$ is much larger than  $\sigma_{\rm Ti} \approx 0.03\mu$s$^{-1}$ at all measured temperatures; thus, the two precession signals in  Eq. (\ref{asym}) were easily separated. Fig.~\ref{TF-2lw}a also shows that $\sigma$  increases sharply below  $T_{\rm c}=18.5$ K due to the increasing field inhomogeneity caused by the superconducting flux lattice, and increases linearly within the statistical errors below about $12$~K. 

\begin{figure}
\centerline{\includegraphics[width=\columnwidth]{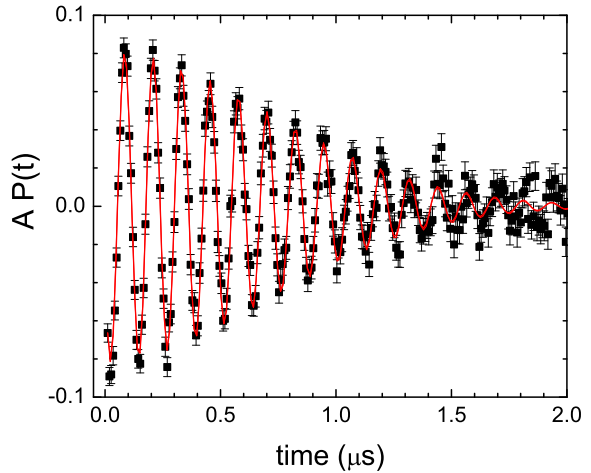}}
\caption{
Muon TF precession signal at $T = 4$~K.  The data from $t = 3 - 10 \mu$s has been subtracted to show  the inhomogeneous relaxation  produced by the flux lattice in the superconducting state. The solid line is a fit for a Gaussian $P(t)$ in Eq. (1).  
}
\label{Gauss-osc}
\end{figure}

A Gaussian time distribution implies a Gaussian field distribution, whereas in a single crystal the field distribution is expected to be asymmetric \cite{Brandt}. The quality of the Gaussian fits is illustrated in Fig. \ref{Gauss-osc} for data taken at $T = 4$K.  Here the $G_{\rm z}(t)$ data (Eq. \ref{asym}) from $t = 3-10\mu$s were fit separately, and this long-time Ti signal was then subtracted from the total spectrum, leaving only the signal from the sample in the superconducting state.  One sees that the Gaussian form for $P(t)$ gives a satisfactory fit.  Thus, the absence of the expected asymmetric field distribution is probably due to the large $\kappa = \lambda(0)/\xi_0 \approx 100$ value, together with the possibility that the flux lattice is distorted somewhat by radiation-induced pinning centers \cite{Brandt}.


We now discuss the data with regard to the superconducting and magnetic properties of PuCoGa$_5$, beginning
with the ZF results.  First, the absence of any temperature dependence below $45$ K in the parameters shown 
in Fig. \ref{ZF-params} is strong evidence that there is no static electronic magnetism. 
The  shape of the observed
short-time Gaussian relaxation shown in Fig.~\ref{ZF} is consistent with static nuclear-dipolar
broadening. (If this relaxation were instead due  to damped, coherent precession from  magnetic order, we estimate an  ordered moment $< 5 \times 10^{-3} \mu_{\rm B}$, which is  $\ll 0.68 \mu_{\rm B}$ obtained from the susceptibility \cite{Sarrao-2002}.) Assuming that muons occupy the same sites as 
in CeRhIn$_5$ \cite{Schenck},  we estimate a ZF nuclear-dipolar linewidth of about $0.29~ \mu$s$^{-1}$ for 
both of the hypothetical $(1/2,1/2,1/2)$ and $(0,1/2,0)$ sites . This  is
about $25$\% smaller than observed, indicating a different site occupancy from CeRhIn$_5$.     Note, however, that none of our conclusions concerning the superconducting state depend at all on knowing the muon site(s).

In TF measurements the $\mu$SR rate from the vortex lattice $\sigma_{\rm v}$ measures the width of the
rms field distribution $\langle\Delta B^2\rangle$:
$\sigma_{\rm v}=\gamma_{\mu}\langle\Delta B^2\rangle^{1/2}$ , where 
$\gamma_{\mu}=8.516\times 10^8$ s$^{-1}$ T$^{-1}$ is the muon gyromagnetic ratio.
The $\langle\Delta B^2\rangle$  is related to $\lambda$
through the expression (assuming a hexagonal vortex lattice and $H_0 \ll H_{\rm c2}$) \cite{Brandt}:
\begin{equation}
\sigma_{\rm v} \propto \langle\Delta B^2\rangle^{1/2}= (0.00371)^{1/2} \Phi_0 / \lambda^2 \propto \eta_{\rm s}/m^*
\label{eq:Brandt}
\end{equation}
where $\Phi_0 = hc/2e = 2.07 \cdot 10^{-7}$~G$\cdot$cm$^2$ is the magnetic flux quantum, $\eta_{\rm s}$ the superfluid density and $m^*$ the in-plane effective mass \cite{tinkham}. 

The temperature dependence of $\sigma_{\rm v}$ is obtained by
subtracting the temperature-averaged normal-state linewidth $\sigma_{\rm n}$ from the $\sigma$ shown
in Fig. \ref{TF-2lw}a:
$\sigma_{\rm v}^2 = \sigma^2 - \sigma_{\rm n}^2$, where $\sigma_{\rm n}$ = 0.20~$\mu$s$^{-1}$. This  assumes that the linewidths $\sigma_{\rm v}$ and $\sigma_{\rm n}$ arise from independent sources, which they do.   To check the effect of the form of $P(t)$ in Eq. (1) on the temperature dependence of the penetration depth, the  time spectra were also fit using $P(t) = \exp(-\Lambda t)$, and 
an analogous subtraction procedure was used (suitable for exponentials), namely $\Lambda_{\rm v} = \Lambda -
\Lambda_{\rm n}$.  As seen in  Fig.~\ref{TF-2lw}b the exponential relaxation rate $\Lambda_{\rm v}$ 
(normalized at the lowest temperature to the Gaussian rate $\sigma_{\rm v}$) yields essentially the same
temperature dependence as does $\sigma_{\rm v}$, with a linear increase below about $12$~K.  

The inset to Fig. \ref{TF-2lw}a shows the temperature dependence of $\lambda(T)$,  extracted from Eq. \ref{eq:Brandt}.    We see that the characteristic length scale for $\lambda$ is $\sim$ 500
lattice constants ($a = b = 4.232$ \AA~\cite{Sarrao-2002}). Thus, if the muon makes $\sim 5$ random hops in the $12 \mu$s 
duration of our experiment
($1/\tau \approx 0.4 \mu$s$^{-1}$) it moves only $\sim \surd5$ lattice spacings, and cannot 
affect the measurement of $\lambda$. We obtain an extrapolated zero-temperature penetration depth $\lambda(0)
= 241$ nm. To obtain the  Ginsburg-Landau penetration depth one must apply mean-free-path
corrections: $\lambda_{GL} = \lambda(0)(1 + \xi_0/l_{\rm tr})^{1/2}$, where $l_{\rm tr}$ is the transport
mean free path \cite{tinkham}.  We have estimated $l_{\rm tr} \approx 10$ nm using the values of the resistivity at $T = T_{\rm c}$, Sommerfeld constant, $T_{\rm c}$ 
and $\xi_0$ \cite{Sarrao-2002} and the formulas found in Ref. \cite{Orlando}, so that $\lambda_{GL} \approx 265$ nm. 
We note that our direct measurement of $\lambda(0)$ is about $2\times$ that reported from critical field data 
\cite{Sarrao-2002}, and  that the magnitude of $\lambda$ can depend on the fit function used \cite{Sonier}.   Thus, our principal result is the linear {\it temperature dependence} of $\eta_{\rm s}$ (Eq. (\ref{eq:Brandt})), or alternatively, the low-temperature linaer behavior of $\lambda(T)$. As shown in Fig. \ref{TF-2lw}a (inset)  $\Delta\lambda(T) = \lambda(T) - \lambda(0) \propto T$ for $T/T_{\rm c} \leq 0.5$, within the statistical errors.

We now discuss the implications of our measurements for the symmetry of the superconducting 
order parameter in PuCoGa$_5$. The ZF $\mu$SR linewidth $\Delta(T)$ shows no change in magnitude below
$T_{\rm c}$. Thus, there is no evidence for a TRV superconducting
order parameter, the signature for which is an increased 
linewidth below $T_{\rm c}$ arising from spin or orbital moments \cite{TRsym}, as found in (U,Th)Be$_{13}$ \cite{Heffner},
Sr$_2$RuO$_4$ \cite{Luke}, and PrOs$_4$Sb$_{12}$ \cite{Aoki}.  This means that the superconducting order parameter is  a linear combination of basis functions for tetragonal symmetry with real coefficients.

In a superconductor whose electrons are paired in an $L = 0$ orbital angular momentum state
the $\mu$SR rate $\sigma_{\rm v}$ or $\Lambda_{\rm v}$ is relatively temperature-independent
below about $T/T_{\rm c} = 0.5$, reflecting exponentially-activated quasiparticle excitations over a superconducting gap which is non-zero over the entire Fermi surface \cite{tinkham}.  This is clearly not observed in PuCoGa$_5$, as seen in 
 Fig. \ref{TF-2lw}b, where the two-fluid approximation \cite{tinkham} for s-wave superconductivity is plotted as the solid curve. Instead, the relaxation rate continues to increase linearly at low temperatures,  yielding a  
  low-temperature T-linear behavior for $\eta_{\rm s}$ and $\Delta\lambda(T<<T_c)$.  
 This behavior is quite generally  associated 
with a line of nodes in the gap function for strong spin-orbit coupling (as expected for the heavy element Pu)\cite{Blount} and  local electrodynamics \cite{sc theory}; e.g., $\Delta\lambda(T) \propto T$  for $T^* < T \ll T_{\rm c}$, where $T^*$ is a cross-over temperature 
$T^* \sim v_{\rm F}/\pi\lambda(0)$, below which $T^3$-dependence is calculated to occur due to non-local effects \cite{non-local}. We estimate the Fermi velocity $v_{\rm F} \propto \xi_0T_{\rm c} \approx 3 \times 10^6$ cm/s \cite{Orlando} and obtain $T^* \approx 1.8$~K. This is consistent with our results.  A linear T-dependence  has been  found in a variety of 
copper-oxide high-temperature  superconductors \cite{Sonier}, and is associated with $d$-wave ($L = 2$) pairing (even parity, pseudo-spin singlet). In many heavy fermion superconductors  $d$-wave pairing has also been established \cite{heffner-norman}.
This particular gap symmetry is strongly enhanced by a spin-fluctuation pairing mechanism \cite{Moriya,spinfluct}, as opposed to 
electron-phonon pairing which favors $s$-wave pairing.  Thus, our ZF and TF $\mu$SR studies in PuCoGa$_5$ show a gap symmetry which is similar to both the heavy fermion and copper-oxides superconductors \cite{Curro}, though PuCoGa$_5$ is neither particularly heavy nor is it an oxide.  This behavior may be tied to the 2-dimensional character of its Fermi surface \cite{PuCoGa5-theory} which can enhance low-frequency spin fluctuations. Future $\mu$SR experiments are planned to investigate the field dependence of the superconducting symmetry of this  material.

Work at LANL and LLNL(contract W-7405-Eng-48)  performed  under
 the U.S. D.O.E. Work
at Riverside  supported by the U.S. NSF, Grant DMR-0102293.
We thank the staff of TRIUMF
 and acknowledge discussions with  N. J. Curro, J.E. Sonier and J. D. Thompson.


\end{document}